\documentclass[aps,prc,twocolumn,nofootinbib]{revtex4}

\usepackage{graphicx}
\usepackage{dcolumn}
\usepackage{bm}
\usepackage[colorlinks=true]{hyperref}

\newcommand{\be}{\begin{equation}}
\newcommand{\ee}{\end{equation}}
\newcommand{\bea}{\begin{eqnarray}}
\newcommand{\eea}{\end{eqnarray}}

\newcommand{\vp}{{\bf p}}
\newcommand{\vv}{{\bf v}}
\newcommand{\gton}{\mathrel{\lower.9ex\hbox{$\stackrel{\displaystyle>}{\sim}$}}}
\newcommand{\lton}{\mathrel{\lower.9ex\hbox{$\stackrel{\displaystyle<}{\sim}$}}}

\begin{document}

\title{Suppression of elliptic flow without viscosity}

\author{Adam Takacs}
\email{adam.takacs@uib.no}
\affiliation{Department of Physics and Technology, University of Bergen, Bergen 5020, Norway}%
\affiliation{Institute of Physics, Eotvos University, Pazmany P. s. 1/A, Budapest 1117, Hungary}%
\affiliation{
Wigner Research Centre for Physics of the H.A.S., P.O. Box, Budapest 1525, Hungary}%

\author{Denes Molnar}
\email{molnar@physics.purdue.edu}
\affiliation{
Department of Physics and Astronomy, Purdue University, West Lafayette, IN 47907, United States
}%
\affiliation{
Wigner Research Centre for Physics of the H.A.S., P.O. Box, Budapest 1525, Hungary}%

\date{\today}

\begin{abstract}
We investigate fluid-to-particle
conversion using the usual Cooper-Frye approach but with more general local equilibrium distributions than the Boltzmann or Bose/Fermi distributions typically used.
Even though we study ideal fluids (i.e., shear stress and bulk pressure are zero everywhere),
we find a suppression of elliptic flow ($v_2$) at high transverse momenta ($p_T\gton1.5$ GeV/c),
relative to results obtained with the traditional Boltzmann distributions. 
The non-viscous suppression shows qualitatively similar
features to the well-known shear viscous suppression of $v_2$;
for example, it increases with $p_T$, and it is smaller for heavier species as seen in self-consistent kinetic theory
calculations.
Our results question whether all of the $v_2$ suppression seen in the data can be attributed to viscous effects,
and indicate that shear viscosities extracted from RHIC and LHC elliptic flow data might be overestimated.
\end{abstract}


\maketitle

\section{Introduction}
\label{sc:intro}

The application of hydrodynamics to model heavy-ion reactions requires
a ``particlization'' model \cite{HuovinenPetersen} to convert the hydrodynamic fields to particles.
The most common procedure in practice is to match locally, 
the fluid variables to a gas of hadrons that is in or near local equilibrium. The matching is made
on a constant energy-density or constant temperature hypersurface (cf. Sec.~\ref{sc:CF}).
For dissipative fluids
this approach is known to be ambiguous \cite{Molnar:2011kx,Molnar:2014fva,Dusling:2009df}
because, for nonzero shear stress, bulk pressure, or heat flow, infinitely many different out-of-equilibrium
hadron phase space densities can describe the same fluid fields. In particular,
for shear corrections, the most popular quadratic in momentum (Grad) distortion of phase space densities 
is not supported by kinetic theory calculations \cite{Dusling:2009df,Molnar:2014fva}. This translates into appreciable systematic uncertainties in
the extraction of transport coefficients from data \cite{Wolff:2016vcm}.

It should be realized, however, that particlization for ideal fluids carries an analogous ambiguity.
At first sight the claim may look impossible because there is a one-to-one mapping between the fluid dynamic variables
and the parameters of the Boltzmann (more precisely, Bose or Fermi\footnote{For brevity we will collectively refer here to the Bose, Fermi, and Boltzmann distributions as ``Boltzmann''.}) phase space distributions of the hadrons in local thermal and chemical equilibrium.  
However, it is not known {\it a priori} that the particle distributions in the hadronic mixture 
created in heavy-ion collisions are Boltzmann or close to Boltzmann. 
First of all, even in local equilibrium, interacting systems have in general non-Boltzmann single-particle distributions \cite{Bagchi:2018}
(see Ref.~\cite{Csernai:1991fs} for an explicit hadronic example).
Moreover, it is not clear whether heavy-ion collisions lead to local thermalization,
or only some steady-state distribution is reached
( because the system is fairly small with short lifetime and expands rapidly).
In fact, far-from-equilibrum evolution can still obey the hydrodynamic equations of motion, provided the system is (nearly) conformal
and locally isotropized in the comoving frame of the fluid \cite{Arnold:2004ti}. (This is because the energy-momentum
tensor in such
isotropic systems has the ideal fluid form $T^{\mu\nu}=\text{diag}(\varepsilon,P,P,P)$ everywhere, with $P\approx \varepsilon/3$ due to conformality.)

Our main purpose here is to investigate the conversion from ideal hydrodynamics to a hadron
gas with non-Boltzmann, but isotropic local equilibrium distributions. In particular, we study how taking Tsallis distributions~\cite{Tsallis1988}
as equilibrium distributions in the hadron gas affects differential elliptic flow.
Earlier works \cite{Osada:2008sw,Tang:2008ud} concentrated on effects on particle momentum spectra.

The Tsallis distribution (or q-exponential) originated from the generalization of statistical mechanics by taking into account finite-size effects. It has been successfully applied to many physical systems, including a statistical description of hadron spectra in high-energy collisions~\cite{BiroG2017, Tang:2008ud}. Here we use the q-exponential distribution to characterize deviations from the local Boltzmann form 
which might occur in finite-size systems that expand rapidly. One could investigate other distributions as well, the general idea would be the same (though results may differ).

\section{Simple four-source model with Tsallis distribution}
\label{sc:4source}

To give some qualitative insight into how replacing Boltzmann distributions would affect differential elliptic flow $v_2(p_T)$, 
we adapt the simple four-source model from Ref.~\cite{Huovinen:2001cy}. 
That model consist of four uniform, non-expanding fireballs of equal volume and temperature, 
boosted symmetrically in back-to-back pairs along the $x$ and $y$ 
axes in the transverse plane with velocities $\pm v_x$ and $\pm v_y$, respectively ($0 < v_y < v_x$). 
The momentum distribution of particles is then given by
\begin{eqnarray}
    f^{(4s)}=f_{v_x}+f_{-v_x} + f_{v_y} +f_{-v_y} \ ,
 \label{eq:f_4source}
\end{eqnarray}
where the subscripts $\pm v_{x,y}$ denote the direction of the boost.
For Boltzmann sources,
\be
f_{{\rm B},\vv} \propto e^{-E_{\rm LR} / T} \ , \quad E_{\rm LR} \equiv p\cdot u = \gamma (E(\vp) - \vv \vp) \ ,
\label{boost_boltzmann}
\ee
where $\vv$ is the boost velocity, 
$\gamma \equiv \sqrt{1-\vv^2}$, 
$u^\mu \equiv \gamma(1,\vv)$ is the four-velocity of fluid flow,
and $E_{\rm LR}$ is the energy of the particle in the comoving frame.

We extend the model by replacing the Boltzmann factor of comoving energy in Eq.~(\ref{boost_boltzmann}) 
with the Tsallis distribution \cite{Tsallis1988} 
\be
f_\alpha(E_{\rm LR})=A\left(1+\frac{\alpha}{\Lambda}E_{\rm LR}\right)^{-\frac{1}{\alpha}}\ ,
\label{eq:tsallis}
\ee
where $\alpha$ and $\Lambda$ are the Tsallis exponent and temperature%
\footnote{In more traditional notation, $q\equiv\alpha+1$ and $T_q\equiv \Lambda$ are used but we prefer to have the Boltzmann limit at zero exponent $\alpha\to0$.}.
In the $\alpha\rightarrow0$ limit, the Tsallis form becomes a Boltzmann distribution with 
temperature $T = \Lambda$. 
But for any $\alpha > 0$, the distribution has a power-law tail $f_{\alpha} \sim p^{-1/\alpha}$ at high energies. 

\begin{figure}[ht]
\leavevmode
    \includegraphics[width=\linewidth]{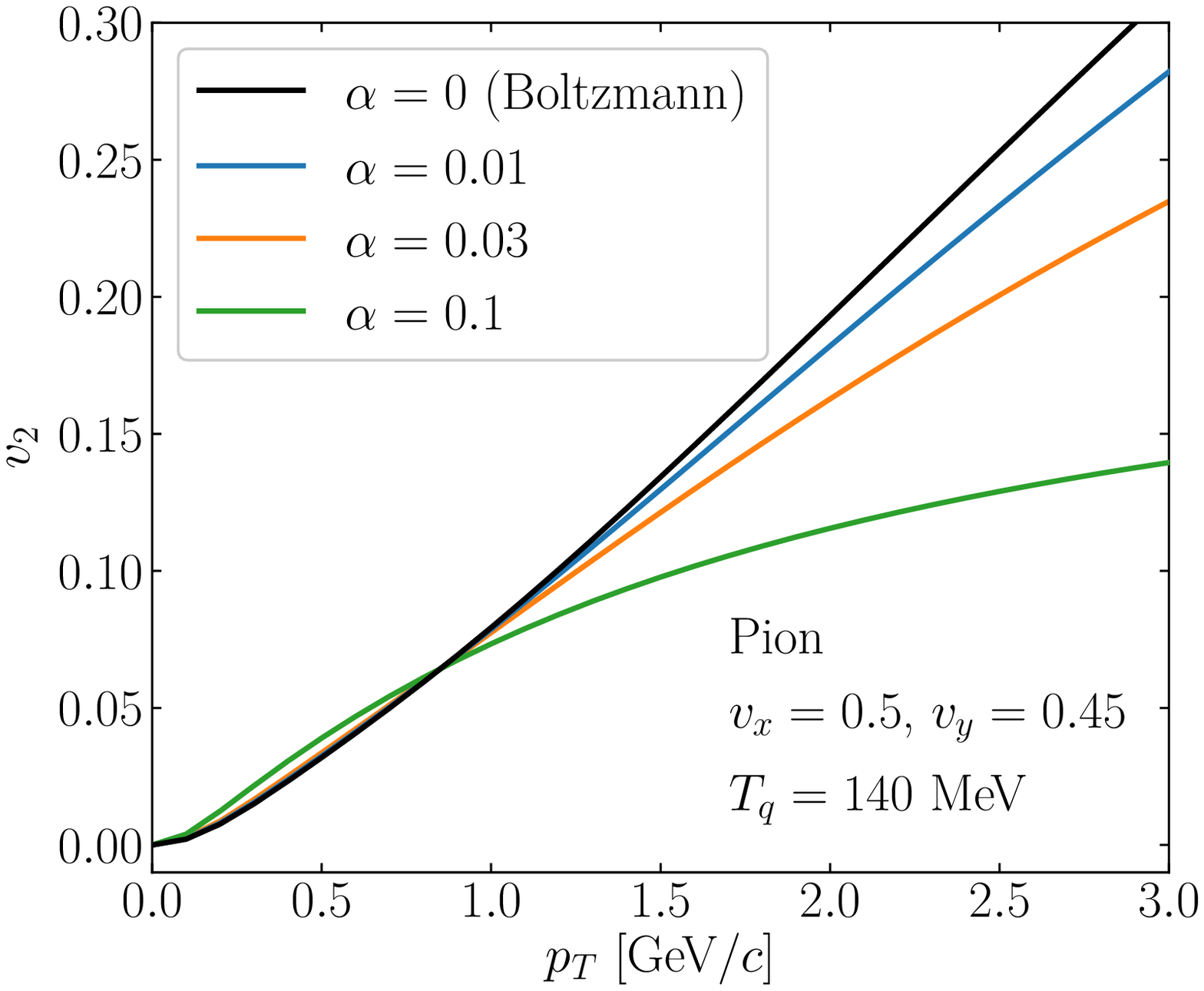}
    \includegraphics[width=\linewidth]{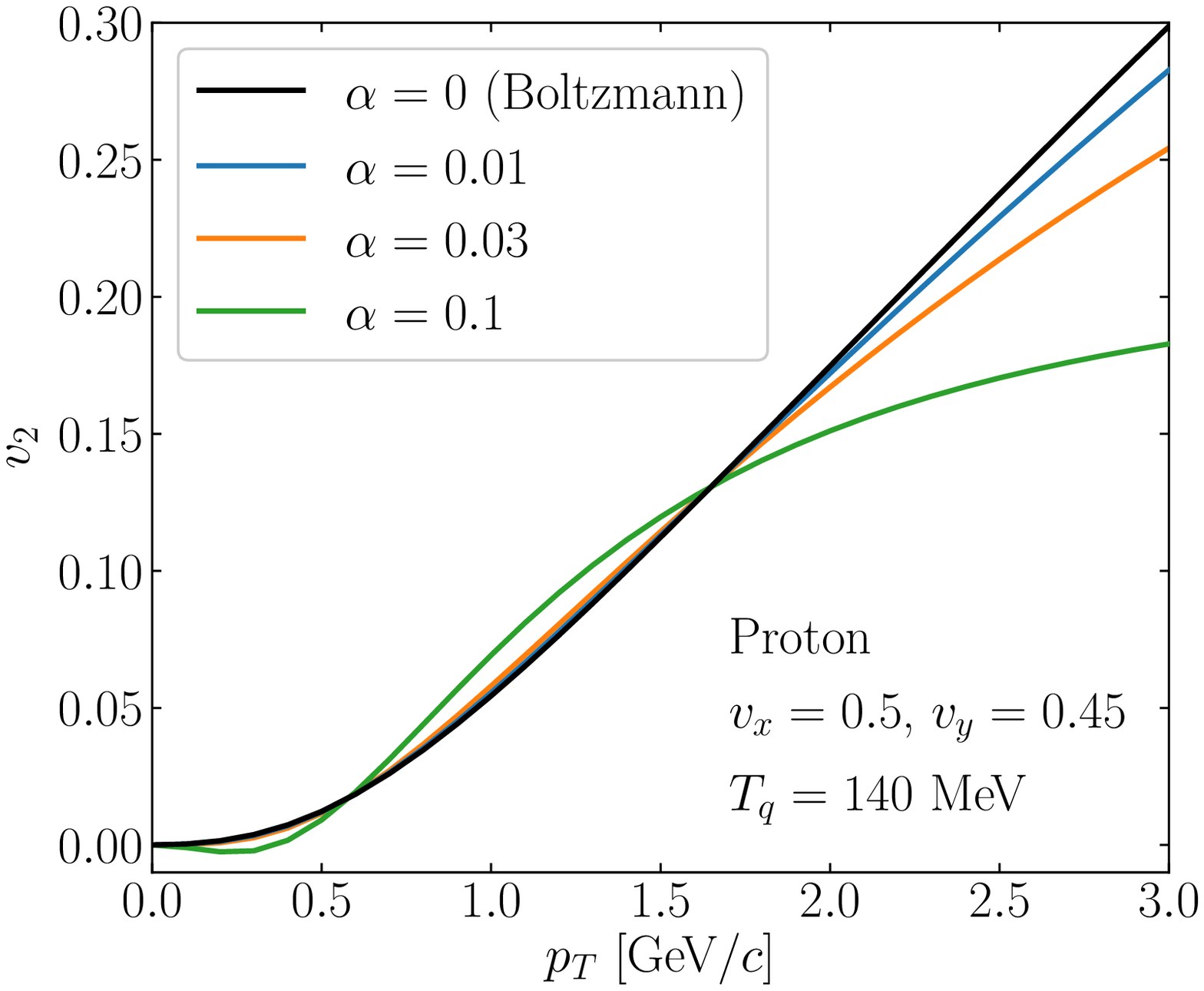}
    \caption{%
Charged pion differential elliptic flow $v_2(p_T)$ from the simple four-source model (Eq.~(\ref{eq:f_4source})) 
for different Tsallis exponents $\alpha$.
Increasing $\alpha$ results in a suppression of anisotropy relative to the Boltzmann case 
($\alpha=0$). Compared to the $v_2$ suppression for pions,
the suppression in proton $v_2$ for Tsallis distributions is smaller.
 }
    \label{fig:4source_all}
\end{figure}

At fixed time, the anisotropic flow coefficient $v_n$ at midrapidity ($y=0$) is, by definition,
\begin{equation}
   v_n(p_T,y=0)=\frac{\int^{2\pi}_0d\phi\,\cos(n\phi)f^{(4s)}(p_T,\phi, y=0)}{\int^{2\pi}_0d\phi\,f^{(4s)}(p_T,\phi,y=0)}.
\label{eq:vn_4source}
\end{equation}
Figure~\ref{fig:4source_all} shows the pion and proton elliptic flow $v_2(p_T)$ resulting from 
Eq.~(\ref{eq:vn_4source}) for different exponents $\alpha$. The source parameters were set to
$v_x=0.5$, $v_y=0.45$, and  $T=140$~MeV, which in the Boltzmann case roughly reproduce the charged hadron 
$v_2$ measured in Au+Au collisions 
at RHIC ($\sqrt{s_{NN}} = 200$~GeV) at $\sim 30$\% centrality.
Compared to the Boltzmann limit (i.e., $\alpha = 0$), 
a positive $\alpha$ leads to a suppression in $v_2$ (and also in $v_n$) that is very similar to the elliptic flow suppression 
due to shear viscosity \cite{Luzum:2008cw,Dusling:2009df,Molnar:2014fva}. 
This is highly remarkable because, in the example here, {\em local shear stress is zero everywhere}.
The suppression comes, instead, from the power-law tails at high momenta.

For Boltzmann distribution, $v_n$ can be analytically calculated \cite{Huovinen:2001cy,Molnar:2014fva} as a ratio of modified Bessel
functions, and one finds $v_n \to 1$ at high $p_T$ (for any even $n$). For $v_2$, this is because
with exponential tails the ratio of yields in the two principal directions, $\phi = 0$ and $\pi/2$,
in the transverse plane is $N(\phi = \pi/2) / N(\phi=0) \sim e^{-ap_T} / e^{-bp_T} \to 1$ as $p_T\to \infty$ 
(there are more particles
at $\phi = 0$, so $a > b$). In contrast, for Tsallis distributions the ratio
$\sim (a p_T)^{-1/\alpha} / (b p_T)^{-1/\alpha}
\to (b/a)^{1/\alpha}$ stays nonzero at high $p_T$, so $v_n < 1$.

Comparing pion and proton $v_2$ in Fig.~\ref{fig:4source_all},
we see at low $p_T$ the well-known mass ordering of $v_2$ (the flow even gets negative for protons).
This is a well-known generic feature of the original four-source model. 
At high $p_T$, we
find a proton-pion difference as well in the suppression of $v_2$ for Tsallis distributions,
which suppression is weaker for protons than for pions. 
This, again, is qualitatively similar to how self-consistent 
shear viscous corrections behave \cite{Molnar:2014fva}, which is a remarkable result from our simple model with no
shear viscosity.

The four-source model presented above is,
of course, not a substitute for real fluid dynamics (this is remedied in Sec.~\ref{sc:results}). 
It does illustrate, however, the general features of 
elliptic flow suppression coming from Tsallis distributions.

\section{Particlization via the Cooper--Frye formalism}
\label{sc:CF}

Next we study how elliptic flow is affected if Tsallis distributions are used in ideal 
fluid dynamical
calculations. The distribution function enters at the end of the calculation at ``particlization'', when
the fluid fields are converted to particles. Most commonly the conversion is performed
instantenously over a three-dimensional (3D) hypersurface in four-dimensional (4D) spacetime,
using the Cooper--Frye formula \cite{CooperFrye}
\be
E\frac{dN_i(x,\vp)}{d^3p} = p^\mu d\sigma_\mu(x) f_i(x,\vp) \ .
\label{cf}
\ee
Here, $d\sigma(x)$ is the local surface element vector that is normal to the hypersurface, $x$ is the spacetime, $p$ is the momentum
and $f_i$ is the phase space density for particle species $i$. Typically a constant energy density
or constant temperature hypersurface is chosen. The hypersurface elements $d\sigma_\mu$ are 
provided by the hydrodynamic simulation in discretized form, numerically, and one then sums contributions
from Eq.~(\ref{cf}) for each element to obtain the momentum distribution of particles.
(For an analytic treatment with Tsallis distributions in simplified geometry, see Ref.~\cite{Urmossy:2009jf}.)

For a one-component ideal fluid, without conserved charges,
the hydrodynamic fields are the components of the energy-momentum tensor
$T^{\mu\nu} = (\varepsilon + P) u^\mu u^\nu - P g^{\mu\nu}$,
where $\varepsilon$ is the local energy density, and $P$ is the pressure.
The requirement that the particle distribution 
should reproduce the local hydrodynamic variables imposes the constraints
\be
    T^{\mu\nu}(x) = \int\frac{d^3 p}{E}p^\mu p^\nu f(x,\vp) \ .
\label{constraints}
\ee
If the momentum dependence in $f$ is through the comoving energy, i.e., if $f$ is isotropic in the comoving frame,
then the constraints fix five parameters%
\footnote{Note that $u$ has only three independent components because$u^2 = 1$.} in $f$:
\bea
\left(
  \matrix{
     \varepsilon \cr
     P \cr
  }
\right) &=& \int \frac{d^3 p}{E} \left(
  \matrix{
    (p\cdot u)^2 \cr
    [(p\cdot u)^2 - m^2]/3 \cr
  }
\right) f \ ,
\nonumber \\
u^\mu &=& \frac{T^{0\mu}}{\sqrt{T^{0\nu} T^0_{\ \nu}}} \ .
\label{matching}
\eea
It is well-known that for Boltzmann phase space density
\be
f_{\rm B}(x,p)=\frac{g}{(2\pi)^3} e^{[\mu(x) - p\cdot u(x)] / T(x)} \ ,
\ee
the constraints Eq.~(\ref{matching}) completely determine all parameters --- namely, 
the local chemical potential $\mu$, temperature $T$, and flow velocity $u^\mu$.

The matching conditions Eq.~(\ref{matching}) can also be used with Tsallis $f_{\alpha}$,
provided we impose $0 \le \alpha < 1/4$ so that all integrals converge.
Specifically,
\bea
\varepsilon &=& 4\pi A \Lambda^4 \int^\infty_z dx\,x^2\sqrt{x^2-z^2}(1+\alpha x)^{-\frac{1}{\alpha}}\ ,
\nonumber \\
P &=& \frac{4\pi A}{3}\Lambda^4 \int^\infty_z dx\,(x^2-z^2)^{\frac{3}{2}}(1+\alpha x)^{-\frac{1}{\alpha}},
\label{matching_Ts}
\eea
where $z=m/\Lambda$. The integrals can be performed numerically,
and one can then invert for the normalization $A$ and the Tsallis temperature $\Lambda$
(we consider the exponent $\alpha$ as a given fix parameter). For example $P/\varepsilon$ gives $z$, i.e., $\Lambda$, 
and then one obtains $A$ from $\varepsilon$.

\begin{figure}[ht]
  \leavevmode
    \includegraphics[width=\linewidth]{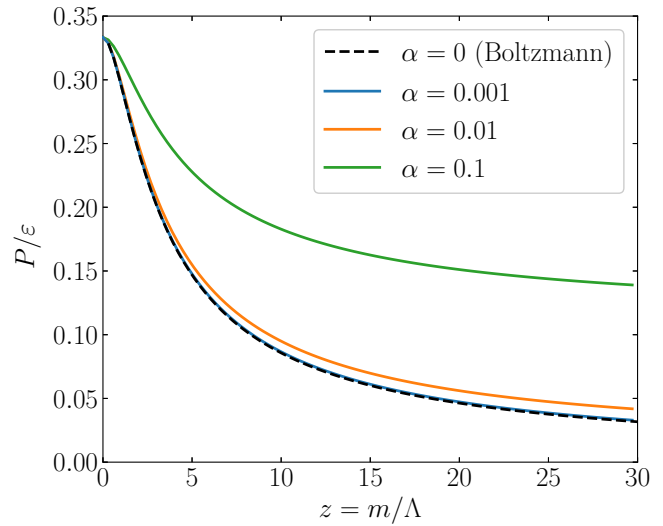}
    \caption{ Pressure to energy density ratio as the function of the inverse temperature for different Tsallis exponents in the range $0 \le \alpha \le 0.1$.
    }
    \label{fig:poe} 
\end{figure}

A minor limitation of the procedure above in the Tsallis case is that, unlike for the Boltzmann distribution,
the pressure to energy density ratio has a nonzero lower bound
$P/\varepsilon \ge \alpha / (1 - \alpha)$.
This is illustrated in Fig.~\ref{fig:poe}, where we plot the ratio versus the dimensionless mass $z$ for different values of $\alpha$. 
At high $z$ (i.e., low ``temperatures''), the ratio does not approach zero, which may make it impossible to match $f$ to the hydro fields
when $P/\varepsilon$ is too low.
In practice, however, this problem has never arisen at the modest $\alpha < 0.06$ values considered in this study.

When applying the Cooper-Frye approach to multicomponent systems, such as a gas of hadrons, 
additional freedom arises because Eq.~(\ref{constraints}) used with $f_i$ gives the partial
contribution to the energy momentum tensor by species $i$, and only the total contribution
$T^{\mu\nu} = \sum_i T^{\mu\nu}_i$ is fixed by the hydro fields. We choose here a simple
prescription where partial pressures $P_i$ and partial energy densities $\varepsilon_i$ are kept the same as in the 
Boltzmann limit, and the matching from Eq.~(\ref{matching_Ts}) is performed to $P_i$ and $\varepsilon_i$ independently
for each species. This means that, in general, each species has its own temperature $\Lambda_i$
and normalization $A_i$. For simplicity, we keep the exponent $\alpha$ the same for all species.
Note that both the local pressure and energy density of the ideal fluid are reproduced exactly,
so shear stress and bulk pressure both remain zero everywhere after conversion to particles.

\section{Results using 2+1D hydrodynamics}
\label{sc:results}

We now apply the approach of the previous Section,
and present results for Au+Au collisions at $\sqrt{s_{NN}} = 200$~GeV
with fluid-to-particles conversion using Tsallis distributions,
on constant temperature $T = T_{\rm conv}$ Cooper-Frye hypersurfaces from relativistic ideal fluid dynamics simulations.
The simulations were performed with the 2+1D {\sf AZHYDRO} code \cite{AZHYDRO}
that solves the ideal fluid dynamic equations with longitudinal boost invariance assumed,
which is a reasonable approximation for observables at midrapidity.
We use a patched version \cite{AZHYDROcode} of the code that includes the {\sf s95-p1} equation of state
parameterization by Huovinen and Petreczky that matches lattice QCD results to a hadron
resonance gas \cite{EOSs95p1}.

For initial conditions at Bjorken proper time $\tau_0 = 0.6$~fm we set
the transverse entropy density distribution $ds/d^2x_T d\eta$ to
a 25\%+75\% weighted sum of binary collision and wounded nucleon profiles,
with diffuse Woods-Saxon nuclear densities for gold nuclei
(Woods-Saxon parameters
$R=6.37$~fm, $\delta=0.54$~fm and we used $\sigma_{NN}^{inel} = 40$~mb).
The baryon density was set to zero everywhere.
The peak value of the entropy density $s_0$ in central collisions ($b=0$) and the temperature of the conversion
hypersurface $T_{\rm conv}$ were set to roughly reproduce pion, kaon, and proton spectra measured by PHENIX \cite{Adler:2003cb},
which gives
\be
    s_0 = 110\text{ fm}^{-3}\ , \quad T_{\rm conv} = 140\text{ MeV}.
\ee
Note that by doing so we ignore subsequent hadron dynamics after particlization, except for resonance decays
(a more sophisticated treatment would evolve hadrons in a transport model, see, e.g., \cite{VISHNU}). 
Simulations were done for three different impact parameters $b=2.5$, $7.3$, $8.7$~fm, which approximately
correspond to centrality classes $0-5$\%, $20-30$\%, and $30-40$\%, respectively.

\begin{figure*}[t]
\leavevmode
    \includegraphics[width=0.329\textwidth]{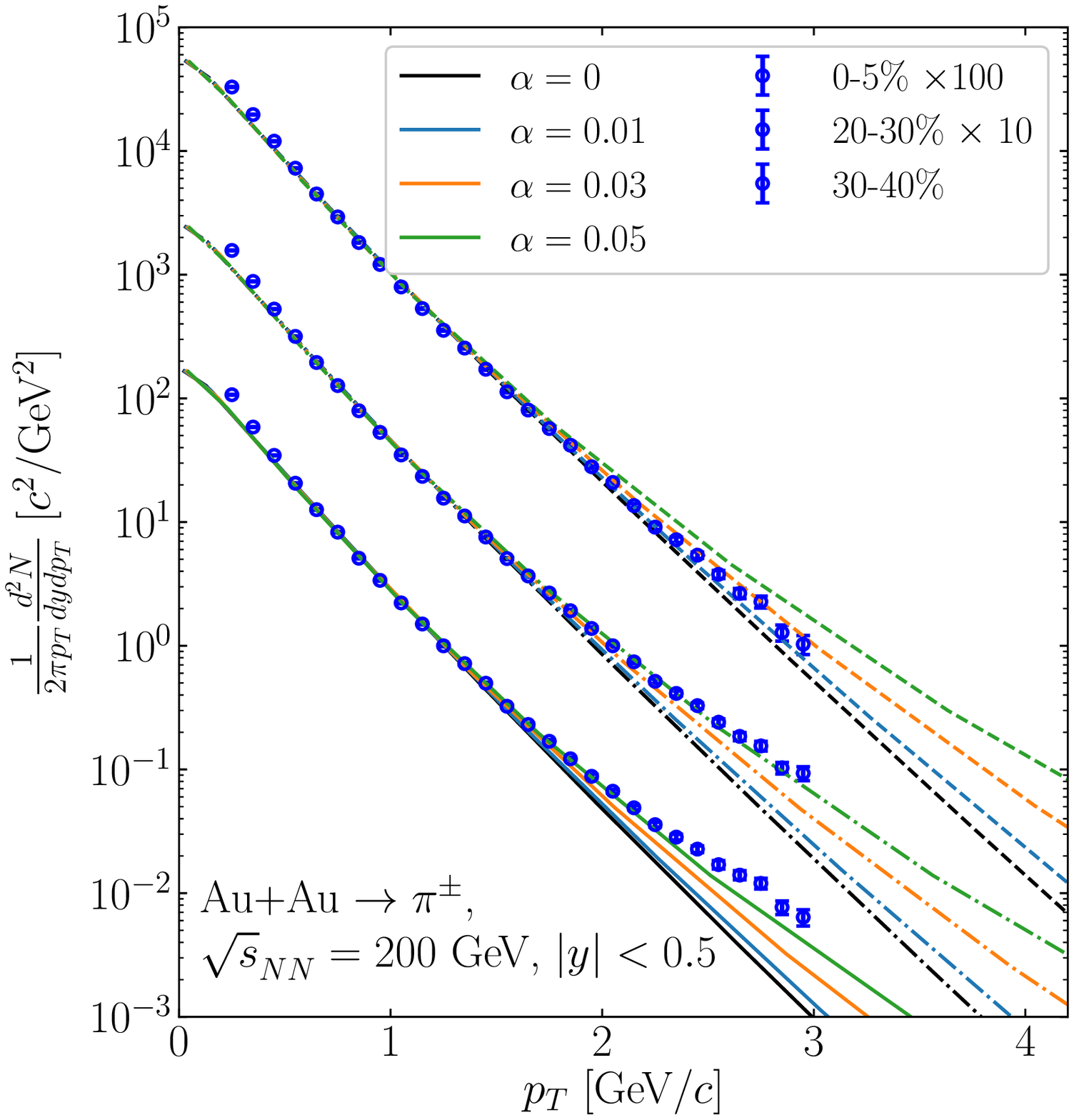}
    \includegraphics[width=0.329\textwidth]{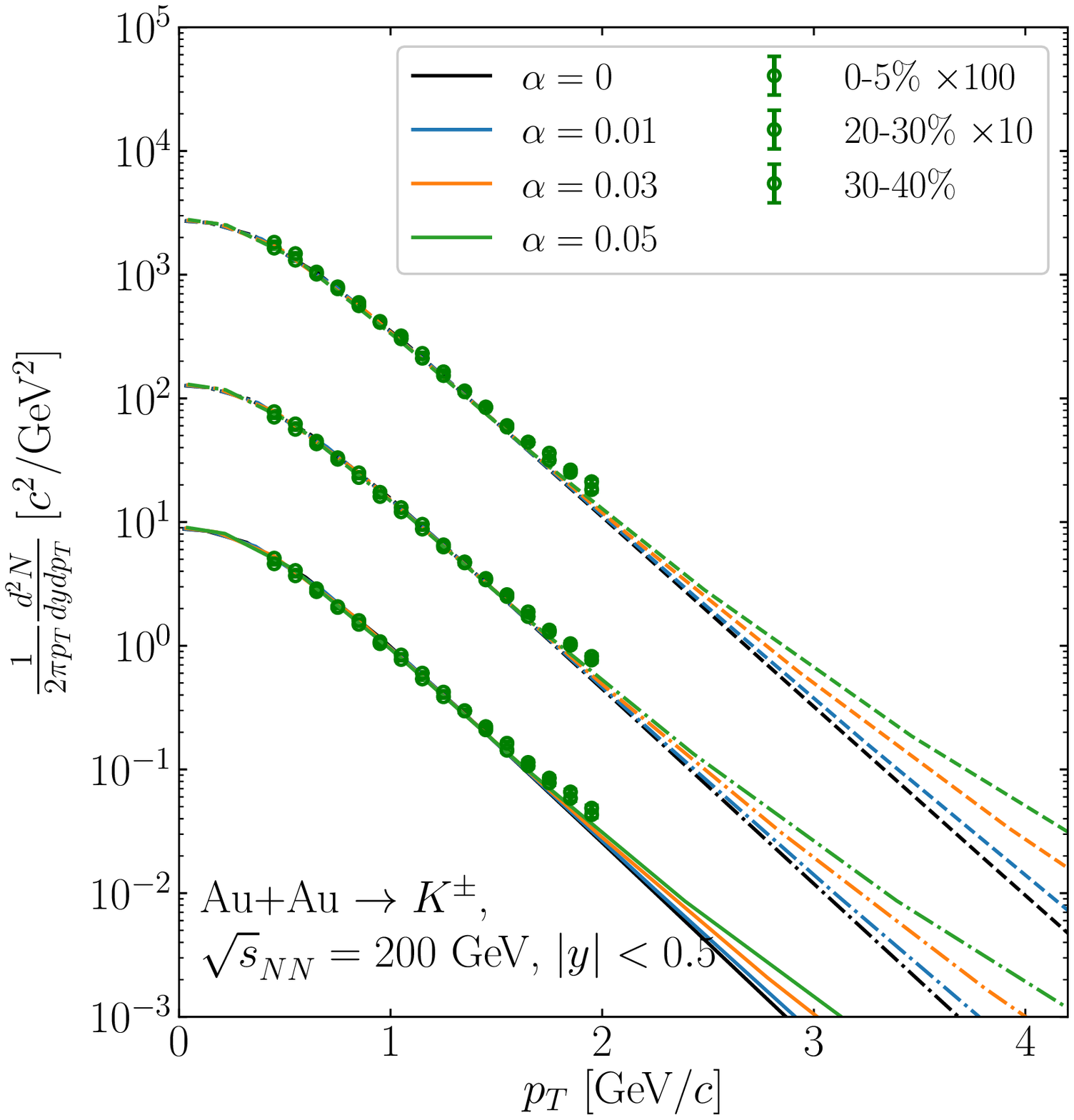}
    \includegraphics[width=0.329\textwidth]{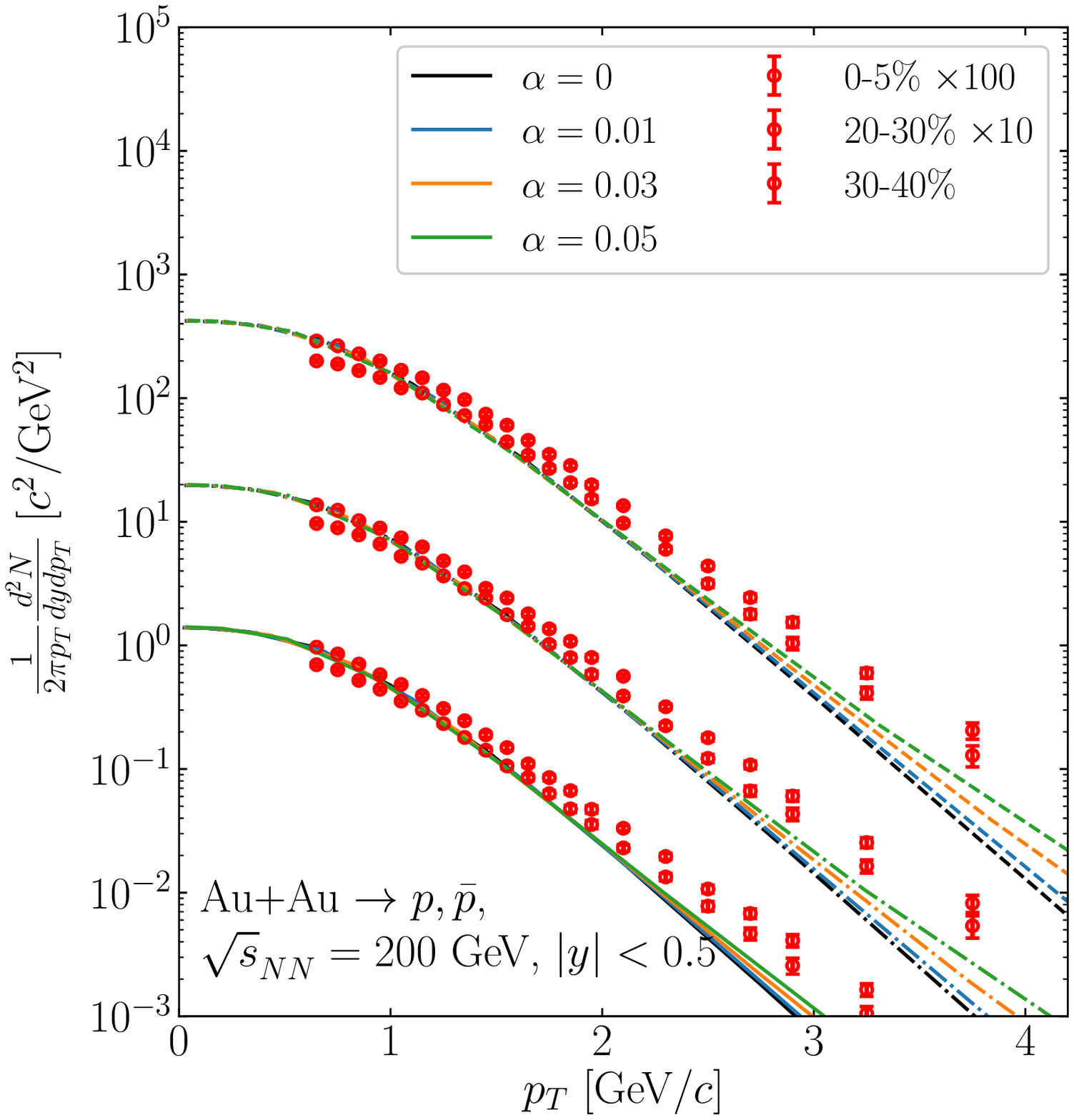}    
    \caption{Identified hadron spectra at midrapidity measured by PHENIX~\citep{Adler:2003cb} in Au+Au at $\sqrt{s_{NN}}=200$~GeV
    for different centralities, compared to spectra from 2+1D ideal hydrodynamic simulations with Tsallis freeze-out with different exponents $\alpha$. The impact parameters are $b=2.3, 7.3$ and $8.7$ fm, corresponding to about 0-5\%,20-30\%, and 30-40\% central events, respectively. Increasing $\alpha$ results in progressively bigger deviations from the Boltzmann case ($\alpha=0$), building up a power-law tail in the spectra that is also seen in experimental data. The different panels are for pion, kaon and protons (from left to right).}
    \label{fig:spectra_all}
\end{figure*}

Fig.~\ref{fig:spectra_all} shows charged pion, charged kaon, and proton spectra in Au+Au at RHIC obtained with Tsallis distributions at the conversion of the fluid to particles.
The $\alpha=0$ curves correspond to the standard Cooper-Frye procedure with Boltzmann distributions.
Three centralities are shown, $0-5$\% (dotted), $20-30$\% (dashed), and $30-40$\% (solid).
For better visibility, spectra for $20-30$\% and $30-40$\% central collisions are shifted upwards
by one and two orders of magnitude, respectively (i.e., factors 10 and 100). 
Feeddown from resonances is included in all results via the {\sf reso} package in {\sf AZHYDRO}.
To guide the eye, experimental data from PHENIX are also shown (filled circles).

Our goal here is not to fit the data perfectly; in fact, agreement is far from excellent.
Still, it is noticeable that the Boltzmann results miss the power-law tails 
seen at $p_T \gton 2$~GeV in the data, while
the use of Tsallis distributions with modest $\alpha \sim 0.03-0.05$ leads to a marked improvement, especially for pions (Fig.~\ref{fig:spectra_all} left panel).
Very similar $\alpha\sim 0.01-005$ values have been 
found in Ref.~\cite{Tang:2008ud} through fits to the Au+Au spectra in a blast-wave model,
while somewhat higher $\alpha \sim 0.08$ was extracted in Ref.~\cite{Osada:2008sw} using a hydrodynamic calculation. For simplicity, we will keep the exponent $\alpha$ the same for all particles and centralities.
But it is clear that the best choice of $\alpha$ varies with centrality and particle species too ---
in fact, at fixed $p_T$, the effect on heavier particles is weaker than on lighter ones similar to hadron spectra fits. For a more detailed experimental study of 
such dependencies, see Ref.~\cite{BiroG2017}.

\begin{figure*}[t]
\leavevmode
    \includegraphics[width=0.329\textwidth]{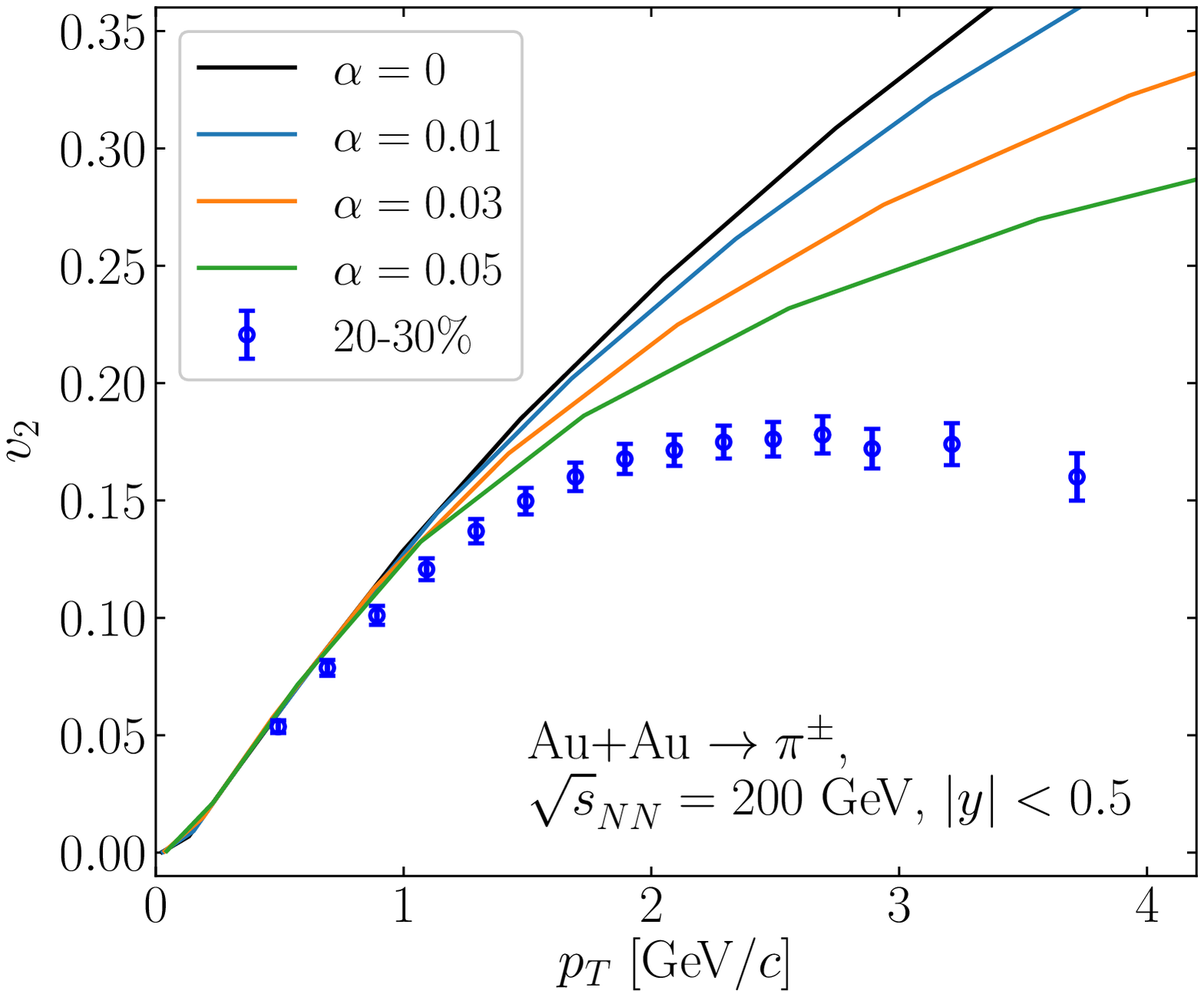}
    \includegraphics[width=0.329\textwidth]{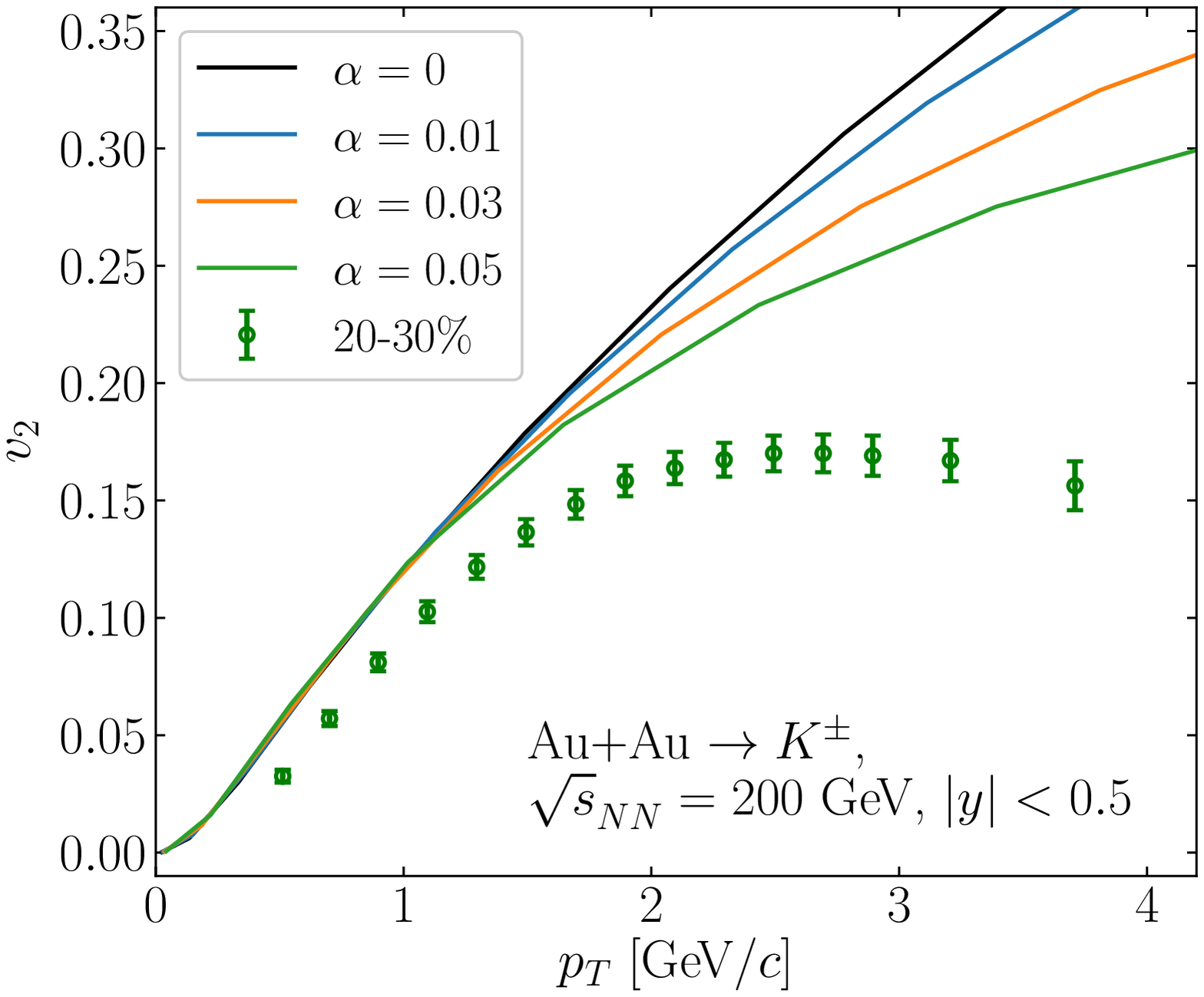}
    \includegraphics[width=0.329\textwidth]{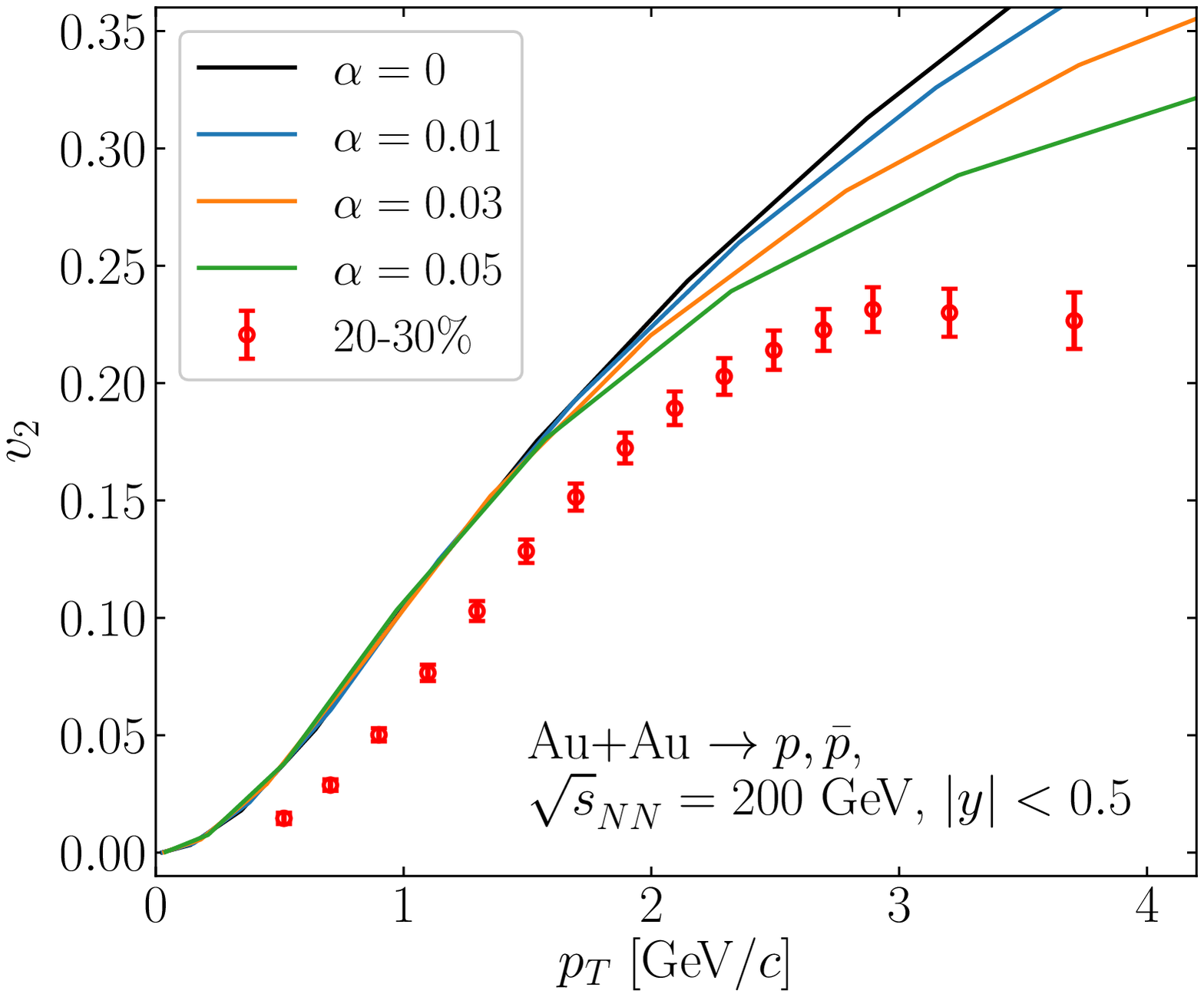}
    \caption{Charged hadron differential elliptic flow at midrapidity in $\sqrt{s_{NN}} = 200$~GeV Au+Au at RHIC with impact parameter $b=7.3$ fm (about 20-30\% centrality), calculated from the same 2+1D ideal hydrodynamic simulations but with different hadron phase space distributions. For increasing $\alpha$, the suppression of $v_2$ relative to the Boltzmann case ($\alpha=0$) becomes stronger. Experimental data from Ref.~\citep{Adler:2003cb} are also shown (filled circles). Panels are for pions, kaons and protons (from the left to the right).} 
\label{fig:v2_all}
\end{figure*}

Finally, Fig.~\ref{fig:v2_all} shows
differential elliptic flow $v_2(p_T)$ for charged pions, kaons, and protons
in Au+Au at RHIC with fluid-to-particle conversion using Tsallis distributions.
We find that all features seen in the four-source model of Sec.~\ref{sc:4source} survive. In particular,
Tsallis distributions give a suppression of $v_2$ at high $p_T$, even for ideal fluids.
The flow suppression progressively increases with the Tsallis exponent $\alpha$,
and at fixed $\alpha$ it is stronger for lighter species.

While the $v_2$ suppression we find from Tsallis distributions is insufficient to explain the experimental data
with absolutely no viscosity, it does move the ideal hydrodynamic results closer to the data.
This raises the question whether all of the elliptic flow suppression is due to viscous effects,
as it is currently assumed by hydrodynamic studies such as \cite{Luzum:2008cw},
or whether only part of the suppression is due to viscosity and 
some of it has other origin, such as non-Boltzmann local equilibrium distributions.

Comparing shear viscous flow suppression calculations in Ref.~\cite{Wolff:2016vcm}, our non-viscous flow suppression for $\alpha\approx0.05$ corresponds to an effective shear viscosity $\eta/s\approx0.05$. There seems to be a very interesting 
connection \cite{Osada:2008sw,BiroMolnar}
between dynamics with non-Boltzmann local equilibria and
dissipative dynamics with Boltzmann equilibria,
which deserves further investigation in the future.

\section{Conclusions}
\label{sec:conc}

In this work we investigated the effect of fluid-to-particle
conversion using the Cooper-Frye approach with Tsallis distributions, 
instead of the usual Boltzman (or Bose/Fermi) distributions.
The main feature we find is a suppression of elliptic flow at high transverse momenta,
relative to the Boltzmann result. Though the suppression shows qualitatively similar
features to the well-known shear viscous suppression of $v_2$,
it appears in our case for ideal fluids, i.e., at zero viscosity.
The main reason for the suppression is the power-law tail of the Tsallis distribution
at high momenta (incidentally, the same power-law tails are known to give a better description of spectra as well).
We first demonstrate the flow suppression in a Tsallis-like generalization of the four-source model of Ref.~\cite{Huovinen:2001cy}
(Sec.~\ref{sc:4source}),
and then show that the suppression is also present for relativistic ideal hydrodynamic simulations of Au+Au collisions at RHIC
(Sec.~\ref{sc:results}).
 
Our results question whether all of the $v_2$ suppression seen in the data can be attributed to viscous effects,
or part of the suppression comes from non-Boltzmann equilibrium distributions. If Tsallis-like equilibrium distributions
play a role, then
there is a possibility that
the shear viscosity extracted from anisotropic flow data at RHIC and the LHC may be overestimated.
It would, therefore, be very interesting to study in the future the interplay between non-Boltzmann equilibrium distributions
and viscous dynamics.

Extensions of our approach to viscous fluids are certainly possible, albeit with significant theoretical
uncertainties because infinitely many such extensions could be postulated.
The situation is analoguous to the state of shear viscous phase space corrections \cite{Molnar:2011kx}, where infinitely many
{\it ad hoc} viscous corrections
functions could be chosen as well. Luckily, kinetic theory
can be used to formulate a self-consistent answer to that problem \cite{Dusling:2009df,Molnar:2014fva}. 
It would, therefore, be very helpful to investigate elliptic flow from 
kinetic theory models that incorporate a non-Boltzmann fixed-point 
distribution \cite{Osada:2008sw,BiroMolnar}.

{\em Acknowledgements.}
Helpful discussions with P.~V\'an, T.~Bir\'o, and P.~Huovinen 
are acknowledged.
AT is supported by the UNKP-18-3 New National Excellence Program of the Hungarian Ministry of Human Capacities and the Trond Mohn Foundation No. BFS2018REK01.
This work was supported in part by the US Department of Energy, Office of Science, 
under Award No. DE-SC0016524, and also by the Wigner GPU Laboratory (DM).



\begin{thebibliography}{99}

\bibitem{HuovinenPetersen}
  P.~Huovinen and H.~Petersen,
  Eur.\ Phys.\ J.\ A {\bf 48}, 171 (2012).

\bibitem{Molnar:2011kx}
  D.~Molnar,
  J.\ Phys.\ G {\bf 38}, 124173 (2011).

\bibitem{Dusling:2009df}
  K.~Dusling, G.~D.~Moore and D.~Teaney,
  Phys.\ Rev.\ C {\bf 81}, 034907 (2010).
  
\bibitem{Molnar:2014fva}
  D.~Molnar and Z.~Wolff,
  Phys.\ Rev.\ C {\bf 95}, 024903 (2017).
  
\bibitem{Wolff:2016vcm} 
  Z.~Wolff and D.~Molnar,
  Phys.\ Rev.\ C {\bf 96}, 044909 (2017).
  
\bibitem{Bagchi:2018} 
  D.~Bagchi and C.~Tsallis,
  J.\ Physa.\ {\bf 491}, 869 (2018).
  
\bibitem{Csernai:1991fs} 
  L.~P.~Csernai, G.~I.~Fai, C.~Gale and E.~Osnes,
  Phys.\ Rev.\ C {\bf 46}, 736 (1992).
  
\bibitem{Arnold:2004ti} 
  P.~B.~Arnold, J.~Lenaghan, G.~D.~Moore and L.~G.~Yaffe,
  Phys.\ Rev.\ Lett.\  {\bf 94}, 072302 (2005).
  
\bibitem{Tsallis1988}
  C.~Tsallis,
  J.\ Stat.\ Phys. {\bf 52}, 479 (1988).
  
\bibitem{Tang:2008ud}
  Z.~Tang, Y.~Xu, L.~Ruan, G.~van~Buren and F.~Wang,
  Phys.\ Rev.\ C {\bf 79} 051901 (2009).
  
\bibitem{Osada:2008sw} 
  T.~Osada and G.~Wilk,
  Phys.\ Rev.\ C {\bf 77}, 044903 (2008).
  Erratum: Phys.\ Rev.\ C {\bf 78}, 069903 (2008).
  
\bibitem{BiroG2017}
  G.~Biro, G.~G.~Barnafoldi, T.~S.~Biro, K.~Urmossy and A.~Takacs,
  Entropy {\bf 19}, 88 (2017).

\bibitem{Huovinen:2001cy}
  P.~Huovinen, P.~F.~Kolb, U.~W.~Heinz, P.~V.~Ruuskanen and S.~A.~Voloshin,
  Phys.\ Lett.\ B {\bf 503}, 58 (2001).

\bibitem{Luzum:2008cw} 
  M.~Luzum and P.~Romatschke,
  Phys.\ Rev.\ C {\bf 78}, 034915 (2008).\\
  Erratum: Phys.\ Rev.\ C {\bf 79}, 039903 (2009).

\bibitem{CooperFrye}
  F.~Cooper and G.~Frye,
  Phys.\ Rev.\ D {\bf 10}, 186 (1974).

\bibitem{Urmossy:2009jf} 
  K.~Urmossy and T.~S.~Biro,
  Phys.\ Lett.\ B {\bf 689}, 14 (2010).

\bibitem{AZHYDRO}
  P.~F.~Kolb, J.~Sollfrank and U.~W.~Heinz,
  Phys.\ Rev.\ C {\bf 62}, 054909 (2000).\\
  P.~F.~Kolb and R.~Rapp,
  Phys.\ Rev.\ C {\bf 67}, 044903 (2003).\\
  P.~F.~Kolb and U.~W.~Heinz,
  *Hwa, R.C. (ed.) et al.: Quark gluon plasma* 634-714.

\bibitem{AZHYDROcode}
The original version 0.2 of
AZHYDRO and version 0.2p2 patched by P.~Huovinen and D.~Molnar
are available on the WWW from the Open Standard Codes and Routines (OSCAR) repository at
http://karman.physics.purdue.edu/OSCAR

\bibitem{EOSs95p1}
  P.~Huovinen and P.~Petreczky,
  Nucl.\ Phys.\  A {\bf 837}, 26 (2010).  

\bibitem{Adler:2003cb} 
  S.~S.~Adler {\it et al.} [PHENIX Collaboration],
  Phys.\ Rev.\ C {\bf 69}, 034909 (2004).

\bibitem{VISHNU}
  H.~Song, S.~A.~Bass, U.~Heinz, T.~Hirano and C.~Shen,
  Phys.\ Rev.\ C {\bf 83}, 054910 (2011).\\
  Erratum: Phys.\ Rev.\ C {\bf 86}, 059903 (2012).
  
\bibitem{BiroMolnar}
  T.~S.~Biro and E.~Molnar,
  Eur.\ Phys.\ J.\ A {\bf 48}, 172 (2012).
 

\end{thebibliography}
\end{document}